**Climate change and shrinking salamanders: Alternative mechanisms for changes in plethodontid salamander body size**

Grant M. Connette[1*], John A. Crawford[2], William E. Peterman[3]

1 – Division of Biological Sciences, University of Missouri, Columbia, MO 65211

2 – Department of Biological Sciences, Lindenwood University, St. Charles, MO 63301

3 – Prairie Research Institute, Illinois Natural History Survey, University of Illinois, Champaign, IL 61820

**Abstract**

An increasing number of studies have demonstrated correlations between climate trends and body size change of organisms. In many cases, climate might be expected to influence body size by altering thermoregulation, energetics or food availability. However, observed body size changes can result from a variety of ecological processes (e.g., growth, selection, population dynamics), yet may also be due to imperfect observation. We used two extensive datasets to evaluate alternative hypotheses for recently reported changes in the observed body size of plethodontid salamanders. We found that mean adult body size of salamanders can be highly sensitive to survey conditions, particularly rainfall, due to the fact that smaller individuals are more likely to be sampled under dry conditions. This systematic bias in the detection of individuals across a range in body size would result in a signature of body size reduction in relation to reported climate trends when it is simply observation error. We also identify considerable variability in body size distributions among years that shows a correspondence with rainfall. This suggests that annual variation in growth or shifting population age structure could also result in a correlation between climate and mean body size. Finally, our study demonstrates that measures of mean adult body size can be highly variable among surveys and that large sample sizes may be required to make reliable inferences. Identifying the effects of climate change is a critical area of research in ecology and conservation. Researchers should be aware that observed body size changes in certain organisms may be a result of either true ecological processes or systematic bias due to non-random sampling of populations. Ultimately, the credibility of ecological research related to climate change depends on researchers demonstrating a thorough consideration of alternative hypotheses for observed changes in species.

**Introduction**

Global climate change has been shown to impact the geographic distributions of species (Hickling *et al.*, 2006, Parmesan & Yohe, 2003, Perry *et al.*, 2005) as well as timing of life cycle events such as breeding and migration (Menzel *et al.*, 2006, Root *et al.*, 2003). Recent studies have also highlighted the potential for climate change to drive body size change in organisms (Baudron *et al.*, 2014, Daufresne *et al.*, 2009, Gardner *et al.*, 2011, Sheridan & Bickford, 2011). Altered thermoregulation, energetics or food availability may ultimately cause climate-driven changes in body size (Gardner *et al.*, 2011), yet observed shifts in typical body size of a population may result from multiple mechanisms. For instance, climate may directly influence size-at-age (e.g., size at maturity) or, alternatively, may cause a shift in population age structure which also leads to a directional change in body size (Daufresne *et al.*, 2009). In the latter case, changes in a number of demographic processes such as recruitment, growth, or survival may be responsible for population body size trends.

Given the imperfect ability of researchers to observe many ecological systems, there is a very real possibility that systematic sampling bias can confound inferences concerning underlying ecological processes such as climate-driven body size change. Sampling bias may arise due to a number of factors such as trap efficiency (Driscoll *et al.*, 2012, Smith *et al.*, 2004, Willson *et al.*, 2008), observer skill (Cunningham *et al.*, 1999, Freckleton *et al.*, 2006, Kéry *et al.*, 2009), habitat characteristics (Peterman & Semlitsch, 2013) and weather conditions (Chandler & King, 2011, O'Donnell *et al.*, 2014, Pellet & Schmidt, 2005). In many of the above cases, sampling bias was related to body size and could lead to larger individuals or species being disproportionately represented in ecological datasets (Cunningham *et al.*, 1999, Freckleton *et al.*, 2006, Smith *et al.*, 2004, Willson *et al.*, 2008). Although relative changes in populations may still be identifiable with consistent bias in sampling, it is also possible that individuals in a population vary in their exposure to sampling due to factors such as their size or survey conditions.

We caution that observed changes in the body size of organisms may result from a number of true ecological processes or systematic bias due to non-random sampling of populations. Here we consider possible hypotheses for recently reported changes in the observed body size of plethodontid salamanders (Caruso *et al.*, 2014). First, it is plausible that climate change has resulted in selection for smaller adult body size in recent decades. Second, individual growth may vary with weather conditions. Third, population age structure may be variable among years (potentially in relation to weather). Fourth, individuals may differ in their exposure to sampling due to survey conditions. In our study, we examine trends in the observed body size distributions of two species of *Plethodon* salamanders using data collected in southwestern North Carolina during separate studies examining the effects of forest management practices on salamander populations. We leverage these extensive datasets to examine the possibility that observed body size distributions are sensitive to proximate survey conditions (rainfall and survey date), to quantify the observed variability in body size among years and to make sample size suggestions for reliable inference concerning mean body size of adult *Plethodon* salamanders.

**Materials and methods**

*Dataset I: Relative abundance of* Plethodon metcalfi

From 2004–2005 we conducted relative abundance surveys for salamanders in 32 (100 x 5 m) sampling plots distributed across 16 sites (2 plots per site) located in the Nantahala National Forest, Macon County, NC, U.S.A. All sites were located between 718 and 1248 m in elevation and were located at least 1 km apart. Ten of these sites were sampled in both 2004 and 2005, while 4 sites were sampled only in 2004 and 2 sites were sampled only in 2005. Each plot was sampled three times during a season (i.e., plots that were sampled in both 2004 and 2005 had a total of 6 visits). We used a nighttime visual encounter search of each plot (survey order was randomized across sites to reduce bias related to seasonal activity) to capture surface-active salamanders. Surveys were performed between 22:00 and 3:00 EST and generally lasted 30 minutes to 2 hours per plot. A researcher walked a straight line through the middle of the plot and searched 2.5 m to the right and left. We identified all captured salamanders to species, weighed and measured for snout-vent length (SVL) and subsequently released all salamanders at the point of capture. We determined age class (adult or juvenile) by comparing the SVL of each individual to published size classes (Bruce, 1967). For the purposes of analyses, we consider 50 mm to be the body size threshold between juveniles and adults, as males of this size often exhibit the secondary sex characteristic of mental gland development. A total of 1,940 *P. metcalfi* were captured across the two field seasons (N = 798 adults).

*Dataset II: Mark-recapture of* Plethodon shermani

From 2009–2013 we conducted capture-mark-recapture surveys for salamanders in 16 (25 x 25m) survey plots located on the Nantahala National Forest, Clay County, NC, U.S.A. These plots were located at similar elevation (~1200 m) in terrestrial habitat and 8 of the 16 plots had timber removed between 2011 and 2013. In this study we consider only data collected pre-harvest or in un-harvested control plots. Surveys were performed between 21:30 and 05:45 EST and generally lasted 1–2 hours per plot. During surveys, we hand-captured all surface-active salamanders encountered during nighttime area-constrained searches of each plot. We individually marked each salamander with visual implant elastomer (e.g., Heemeyer *et al.*, 2007) and recorded its sex, mass and SVL prior to returning them to within 5 m of their unique capture location, almost always on the second night after capture.

In five years, we visited each plot 11–13 times and recorded 13,816 total captures of 9 salamander species. Red-legged salamanders (*P. shermani*) represented the majority of captures, though some individuals showed morphological evidence of hybridization with *P. teyahalee* (e.g., Walls, 2009). We thinned our dataset to 10,187 *P. shermani* individuals (N = 3,758 adults) to exclude captures that occurred after timber harvest. Except for analyses which directly estimate individual age class, we consider 45mm to be the body size threshold between juveniles and adults, as males of this size often exhibit the secondary sex characteristic of mental gland development.

*Mixed effects modeling*

To determine the effects of local climate and season on mean adult SVL and number of adult salamanders counted per survey, we used linear mixed effects models with Gaussian and Poisson error distributions, respectively. For each analysis we conducted model selection on six *a priori* models (Appendix S1), which included days since a soaking rain event and Julian day as fixed effects, each scaled and centered. A soaking rain event was when ≥ 5mm of rain fell within a 24-hour period, as this amount of rainfall is sufficient to reach the forest floor and moisten leaf litter (O'Connor *et al.*, 2006). This measure has also been found to significantly relate to the detection of plethodontid salamanders (Peterman & Semlitsch, 2013). We determined the mean SVL and number of adult *P. metcalfi* collected on a given survey night (N = 42 nights), and used these measures as our response variables to examine the effects of rainfall and survey date on observed body size and salamander counts. Year was considered as a random effect. For *P. shermani*, we calculated the mean SVL and number of adults collected on each survey plot on each survey night (N = 162 surveys). We modeled Julian day nested within survey plot, nested within year as random effects to account for multiple sites being surveyed on a given night under similar conditions, for the repeated sampling of sites, and for multiple years of observations. All mixed effects models were constructed using *lme4* (Bates *et al.*, 2013) and model selection was done using $AIC_c$ as calculated in *AICcmodavg* (Mazerolle, 2012).

*Mixture Analysis*

We also used our 5-year dataset for *P. shermani* to examine among-year variation in the observed body size distribution of this species. Each year, the overall size distribution of captured individuals showed three separate peaks (Fig. 1). Based on growth data from 2,046 recaptured individuals, these three distributions are known to correspond to distinct age classes. The smallest size distribution is composed of hatchling individuals, which are available to sample for the first time in the current season. The middle distribution consists of 2nd-year individuals and overlaps slightly with the largest body size distribution of individuals 3 years and older. Individuals in the largest size class are considered adults, as even male salamanders on the smaller end of this size class will often show mental gland development. We performed a Bayesian analysis of a normal mixture model to identify shifts in the mean body size of the adult (3+ years) age class among years. A basic model represents the overall body size distribution of individuals as the weighted sum of three Gaussian densities.

$$S_i \sim \sum_{c=1}^{3} \omega_c Normal(\mu_c, \sigma_c)$$

Here, body size (S) of each individual $i$ is distributed according to a mixture of three normal probability density functions, where $\omega_c$ represents the probability that any individual belongs in a given body size class, $c$. Thus, our analysis treats the size class membership of each individual as an unobserved (latent) variable that is estimated directly from the data. The parameters $\mu_c$ and $\sigma_c$ represent the mean and standard deviation describing the distribution for each body size class, $c$. In our analysis, we treat year as a fixed effect and separately estimate the parameters $\omega_c$, $\mu_c$ and $\sigma_c$ for each of our five survey years. We can then easily compute the estimated pairwise differences in mean adult body size among years, along with corresponding credible intervals (CRIs). See supporting information (S2) for additional model details, prior specification and R code for this analysis.

*Data Resampling*

To determine the sampling effort required to have confidence in mean adult body size estimates of *P. metcalfi* and *P. shermani*, we repeatedly pooled individual body size measurements from randomly selected surveys and calculated mean adult body size for each new sample of individuals. We varied the number of surveys contributing to the new sample of body size measurements from 1 to 50 and performed 10,000 resamplings for each hypothetical number of surveys. We then calculated the mean percent difference of these samples from the observed mean adult body size based on all original surveys.

**Results**

Mean adult body size of *P. metcalfi* was highly sensitive to survey conditions and was best predicted by a model including rainfall, Julian date and a quadratic term for Julian date (Table 1). This result suggests that large adults were disproportionately represented in samples collected shortly after rainfall and towards the middle of the summer active season (Fig. 2). The effects of rainfall and Julian date were considerable in this study; mean body size of a sample was predicted to vary by up to 11.3% across the observed range of rainfall and by up to 8.3% in relation to Julian date. This equated to a predicted 11.8% difference in mean body size across the range of observed sampling conditions. In our second species, *P. shermani*, we observed no relationship between adult body size and either rainfall or Julian date (Table 1). The number of adult *P. metcalfi* observed was also best explained by rainfall and Julian date, with counts predicted to decrease with time since rainfall and toward the middle of the summer (Table 1, Fig. 2). Counts of adult *P. shermani* also decreased with time since a soaking rain (Table 1, Fig. 3).

Using our *P. shermani* dataset, we found evidence for large annual variation in mean body size of our three age classes. In particular, we note the high variability in both the relative frequencies and body size distributions of the hatchling and juvenile age classes, suggesting substantial annual variation in growth and recruitment rates. Although the adult size class appeared to show the least variability among years, we still observed a 7.0% increase (95% CRI; 4.6–9.3) in mean adult body size between 2009 and 2013 (Fig. 4). Over this period, the estimated mean for adult body size increased from 52.2 to 55.9 mm (SVL). In addition, the difference in mean adult body size averaged 3.6% (95% CRI; 2.6–4.6) among sequential years. Mean adult body size was also higher in years with greater cumulative rainfall over the study period (Fig.5).

Due to the variation in mean adult body size among surveys (Fig. 6 a,b), our data resampling exercise indicates that repeated surveys are required to obtain consistently unbiased estimates of mean adult body size (Fig. 6 c,d). The mean number of adults captured per survey was 19.0 ± 19.8 (SD) for *P. metcalfi* and 26.4 ± 16.7 (SD) for *P. shermani*. Based on all captures, we observed a mean adult body size of 58.6 ± 6.3 mm (SD) for *P. metcalfi* and 54.1 ± 5.4 mm (SD) for *P. shermani*. With just one survey of a population, the survey mean would be expected to differ from the overall mean by an average of 3.96% in *P. metcalfi* and by 2.25% in *P. shermani*. In addition, there is a considerable probability of even greater bias in individual sample means as the 95% confidence region extends to 12.4% and 6.6% in *P. metcalfi* and *P. shermani*, respectively. Random selection of 7 or more surveys, averaging 19.0 individuals, was required to obtain an expected bias of <1% in *P. metcalfi* (~ 133 individuals). For *P. shermani*, 5 random surveys, averaging 26.4 individuals, resulted in an expected bias of <1% (~ 132 individuals). We also observe that, in both species, the 95% confidence region still extended to greater than 3% at this threshold.

**Discussion**

Our study demonstrates that observed body size distributions can be greatly influenced by both sampling bias and relatively short-term population cycles. Specifically, we found that large adults had a disproportionately high probability of being collected shortly after rainfall, that surface activity varied throughout an active season, that there can be considerable annual variability in body size distributions and that relatively large sample sizes and repeated surveys are required to make reliable inferences concerning mean adult body size for both of our salamander species. Not only do sampling variation and fluctuating population size distributions provide a complex background from which to isolate long-term trends, but the systematic effects of weather on sampling and population dynamics can potentially result in emergent patterns over decades, which might be misconstrued as directional selection.

Although imperfect observation is a reality of ecological studies across taxa, plethodontid salamanders may be particularly susceptible due to their highly fossorial nature and sensitivity to temperature and moisture conditions. These salamanders extensively use below-ground retreats or cover objects (Grover, 2006, Petranka & Murray, 2001, Taub, 1961) and detectability of many species is related to rainfall (Connette & Semlitsch, 2013, Peterman & Semlitsch, 2013, Petranka & Murray, 2001). This can be a serious concern when using raw counts of individuals to assess trends in population growth through time (Grant, *In Review*). Our current study provides additional evidence that counts of salamanders are related to rainfall, but also vary temporally throughout the summer active season. Furthermore, we found that individuals within a population may not be equally available for sampling under certain survey conditions. Mean adult body size of *P. metcalfi* was strongly predicted by survey date and time since rainfall. There was no relationship between mean adult body size of *P. shermani* and either rainfall or survey date, possibly because conditions are more uniformly favorable for ground surface activity at higher elevations. Although our results suggest differences may exist among species, the possibility for individuals of differing size to be systematically over- or under-represented in samples due to rainfall suggests that long-term patterns in rainfall may also generate parallel trends in observed body size that are simply due to non-random sampling of individuals.

High variability in mean adult body size across surveys poses an additional problem for studies of body size change. Based on the two species we considered, it is expected that mean body size of salamanders based on small numbers of surveys will be substantially biased relative to the mean from a much larger sample (Fig. 6). In addition, a slight majority of surveys for both species (59.5% and 54.5%, *P. metcalfi* and *P. shermani*) yielded below-average body size measures due to the fact that surveys conducted shortly after rainfall often sampled large numbers of individuals that were of relatively larger size. Based on our data, we would provide a general recommendation that at least 130 body size measurements from across 5-7 temporally spaced surveys be collected to generate point estimates for mean body size of *Plethodon* salamanders. Such a sampling effort aims to achieve an expected bias of <1% relative to the mean from more intensive sampling, with 95% confidence that such a sample will deviate by no more than approximately 4%. We note that large, pervasive body size declines are not apparent in Caruso et al. (2014) when we consider only species that had > 130 total measurements (N=9). For this calculation, we used the estimated annual change in SVL from Table 1 of Caruso et al. (2014; "Slope" values) to recover the expected body size change over 55 years for each species and found an average change of just -0.57% relative to a baseline of the mean size reported for the first decade of sampling. In addition, there was a clear tendency for percent body size change to be greatest for species with fewer individual measurements and populations considered (Fig.

7). Such sample size issues have also been demonstrated by Adams & Church (2008) to be a likely cause for spurious identification of latitudinal clines in amphibian body size (i.e., Bergmann's rule).

      Finally, our study demonstrates that individual growth and short-term demographic changes can be responsible for observed trends in body size. We observed complete shifts in the size ranges of the three *P. shermani* age classes among years (Fig.1), whereas sampling bias would only be expected to influence the relative capture frequencies of individuals within a size distribution. For instance, the mean size of hatchlings, juveniles and adults were all considerably larger in 2013, a particularly wet year, than in 2012 (+22.8%, +15.1% and +4.8%, respectively; Fig. 1). Although we have just five years to consider, there was a clear correspondence between the amount of rainfall over the active season and the mean body size of adults (Fig. 5). The 7% increase in adult body size of *P. shermani* from 2009–2013 suggests that rapid body size change can be observed in less than a generation, likely due primarily to a flexible growth response of salamanders to weather conditions (e.g., Bendik & Gluesenkamp, 2013). The strength of this effect over a five-year period would also indicate that observed trends in body size over relatively few generations are not likely driven by a genetic shift in size at maturity. This is evidenced by the fact that size did not change continuously over the five year period, but fluctuated up and down in close correspondence with active season rainfall. We also found that the relative proportions of hatchling, juvenile, and adult individuals changed substantially through time (Fig. 1). The timing and influence of such population cycles on long-term patterns of body size are difficult to predict, but it is possible that years of high reproductive success and/or juvenile survival could subsequently result in smaller mean adult body size as large numbers of juveniles are recruited into the adult age class. Thus, it is important to recognize that

a reduction in mean body size may be due to an increased frequency of small individuals (a possible indicator of increasingly healthy, productive populations) or a change in the size and/or frequency of large individuals.

Temperature is of critical importance for ectothermic organisms, and climate warming is predicted to affect body size through numerous interacting pathways (Ohlberger 2013). However, climate change is also a spatially and temporally complex phenomenon, and variation in local climate conditions due to landscape topography may be considerable (Dobrowski, 2011, Sears *et al.*, 2011). As a result, species may be able to behaviorally mitigate the effects of climate change, to a certain extent, by taking advantage of favorable local or microclimatic conditions (Sears *et al.*, 2011). Identification of mechanisms for body size shifts in populations may be further complicated by the fact that climate change more generally entails changes in the periodicity and intensity of precipitation events (Kundzewicz *et al.*, 2007). Our results, and those of Bendik and Gluesenkamp (2013), suggest that precipitation may be a fundamental driver of ground surface activity, population dynamics, and growth rates in plethodontid salamanders. Milanovich et al. (2006) also found a relationship between clutch size and annual precipitation, further highlighting the potential contribution of precipitation to population size structure. Thus, it is important to identify whether changes in body size are directly related to temperature as opposed to change in other environmental conditions (Gardner *et al.*, 2011).

Climate change has the potential to profoundly impact species, either directly by altering thermodynamics and the energetic cost of routine daily activities (Ohlberger, 2013) or indirectly by modifying food availability, predator-prey dynamics and community composition (Durant *et al.*, 2007). Although such factors have been previously reported to influence body size distributions of populations, researchers should be aware that observed body size change can result from a variety of ecological processes (e.g., growth, selection, population dynamics), yet may also be due to imperfect observation. In particular, when individuals in a population vary in their availability for sampling under certain weather conditions, there is a high probability that observed population trends relative to climate will be substantially, if not entirely, driven by systematic bias in the sampling or observation process itself. Based on data we present in this study and the extensive literature demonstrating sampling issues in plethodontid salamanders (Bailey *et al.*, 2004a, Bailey *et al.*, 2004b, Bailey *et al.*, 2004c, Bailey *et al.*, 2004d, Buderman & Liebgold, 2012, Connette & Semlitsch, 2013, Dodd & Dorazio, 2004, Hyde, 2001, Peterman & Semlitsch, 2013, Petranka & Murray, 2001), we believe that the correspondence of both natural ecological processes and sampling biases with rainfall have high potential to be misidentified as long-term trends of body size reduction.

The risks of conducting scientific inquiry under a single hypothesis, as opposed to a suite of competing hypotheses, have been clearly established for more than a century (Chamberlin, 1965). We would encourage researchers to objectively evaluate multiple hypotheses for observed changes in species (e.g., Daufresne et al., 2009), which includes acknowledging and accounting for the fact that contributing mechanisms may be statistical or procedural in nature (Grant, *In Review*). Understanding the underlying causes of body size change is an important step toward appropriately directing species conservation and management efforts (Ohlberger 2013). Because such management actions should ideally be rapid and decisive, a misguided sense of certainty concerning existing threats could be detrimental to future management and conservation efforts due to a misdirection of resources towards problems that do not truly exist.

**Acknowledgements**

We thank J. Alexander, T. Altnether, K. LaJeunesse Connette, K. Corbett, B. Cosentino, K. Dipple, D. Hocking, J. Lewis, M. Osbourn, W. Overton, K. Pursel and S. Schlick for assistance with fieldwork.  E. H. C. Grant and members of the Semlitsch lab provided valuable feedback. We also thank the staff of the Highlands Biological Station and the U.S. Forest Service for accommodating this research.  G.M.C. and J.A.C. conducted fieldwork with support from the Highlands Biological Foundation. G.M.C. was also supported by a University of Missouri Life Sciences Fellowship. J.A.C. received additional support from a GAANN fellowship from the U.S. Department of Education and a cooperative agreement with the U.S. Forest Service Southern Research Station. In addition, this publication was developed under a STAR Research Assistance Agreement No. FP917444 awarded by the U.S. Environmental Protection Agency.  It has not been formally reviewed by EPA.  The views expressed in this document are solely those of the authors and EPA does not endorse any products or commercial services mentioned in this publication.  Research was conducted under North Carolina Wildlife Resources Commission collection permits (1137 and SC00405) and approved IACUC protocols through the University of Missouri (3951 and 6144) and the Highlands Biological Station.

**Supporting information**

Appendix S1. *A priori* models considered as predictors of mean adult body size (SVL) and counts in linear mixed models

Appendix S2. Additional details of the normal mixture analysis, prior specification and R code.

Table 1. Ranking of models for mean adult body size and number of adults encountered

| Adult Body Size | K | ΔAICc | ω | Cum. Wt. |
|---|---|---|---|---|
| *P. metcalfi* | | | | |
|     Global* | 6 | 0.00 | 0.89 | 0.89 |
|     Rain + Date | 5 | 5.21 | 0.07 | 0.96 |
|     Rain | 4 | 6.96 | 0.03 | 0.99 |
|     Date$^2$ | 5 | 9.61 | 0.01 | 0.99 |
| | | | | |
| *P. shermani* | | | | |
|     Null | 6 | 0.00 | 0.32 | 0.32 |
|     Rain | 5 | 0.13 | 0.30 | 0.63 |
|     Rain + Date | 7 | 2.06 | 0.12 | 0.74 |
|     Date | 6 | 2.16 | 0.11 | 0.85 |
|     Date$^2$ | 7 | 2.18 | 0.11 | 0.96 |
|     Global* | 8 | 4.27 | 0.04 | 1.00 |
| | | | | |
| **Number of Adults** | K | ΔAICc | ω | Cum. Wt. |
| *P. metcalfi* | | | | |
|     Global* | 5 | 0.00 | 1.00 | 1.00 |
| | | | | |
| *P. shermani* | | | | |
|     Rain | 5 | 0.00 | 0.50 | 0.50 |
|     Rain + Date | 6 | 0.77 | 0.34 | 0.84 |
|     Global* | 7 | 2.46 | 0.15 | 0.99 |
|     Null | 4 | 8.47 | 0.01 | 0.99 |

ΔAICc represents the difference in AICc value between each model and the best model in the set. ω gives the Akaike weight for each model. Table includes only models with ω > 0.
* Global models include Rain (days since ≥ 5mm), Date (Julian day), and Date$^2$

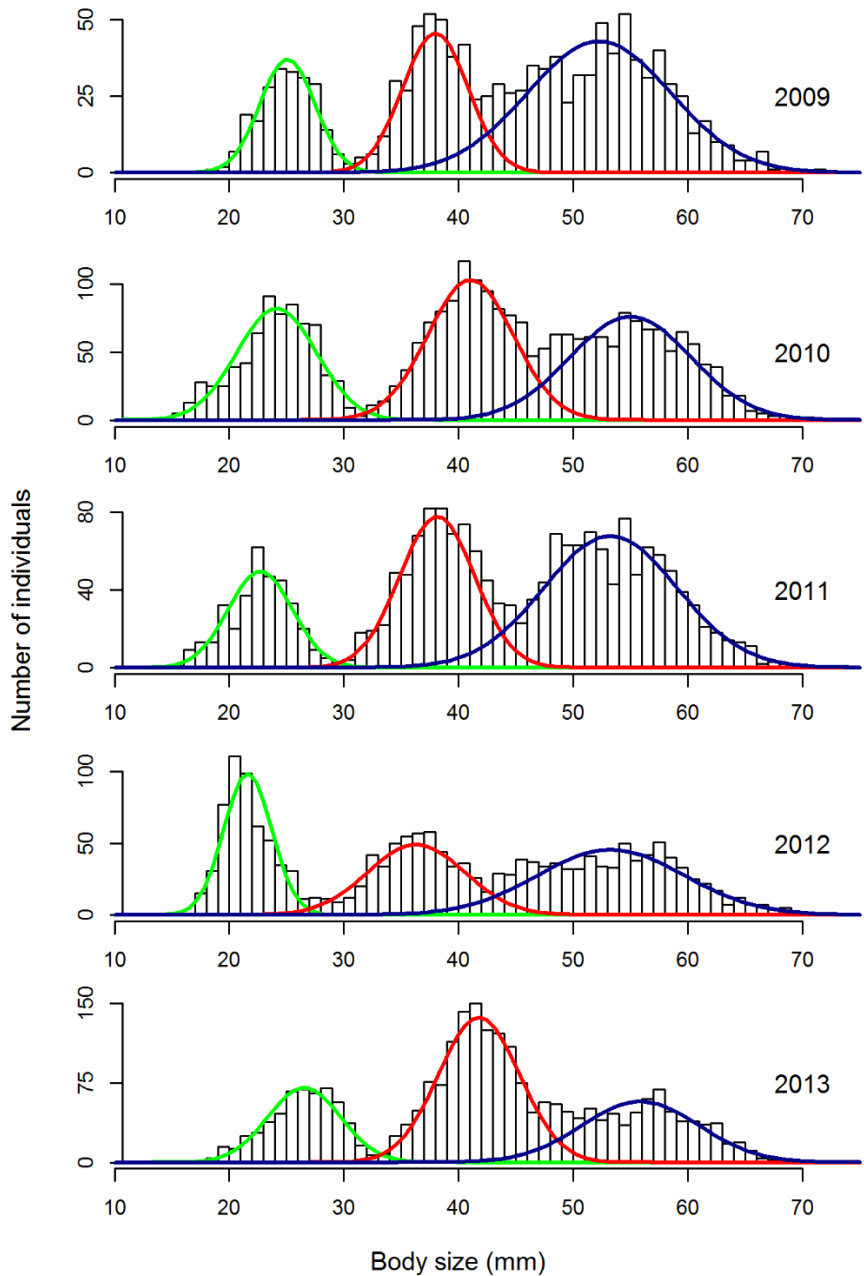

Figure 1. Histograms depict observed body size distributions of *P. shermani* from 2009 (top panel) through 2013 (bottom panel). Lines indicate estimated densities by size class based on analysis of a normal mixture model.

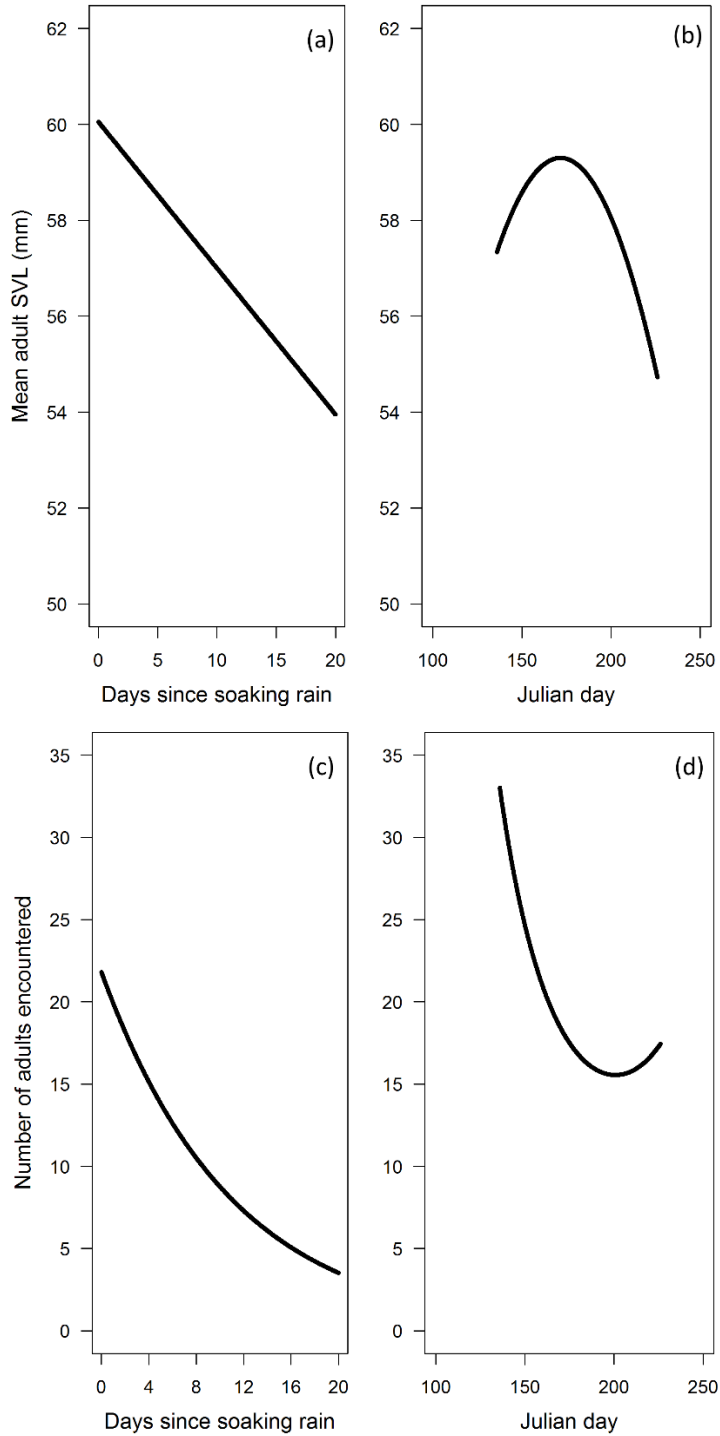

Figure 2. Estimated effects of time since a soaking rain (a) and Julian day (b) on mean adult SVL of *P. metcalfi*. Estimated effects of time since soaking rain (c) and Julian day (d) on the number of adults encountered in a survey.

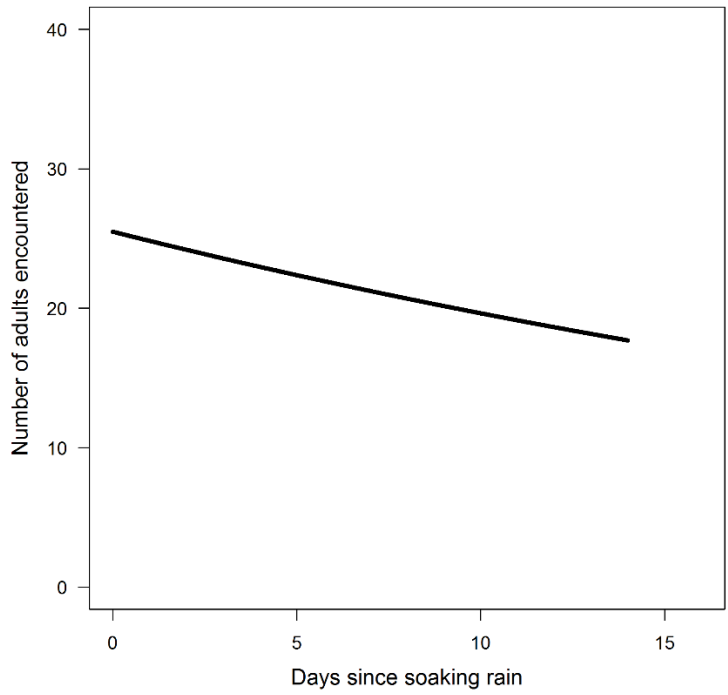

Figure 3. Estimated effects of time since a soaking rain on the number of adult *P. shermani* encountered.

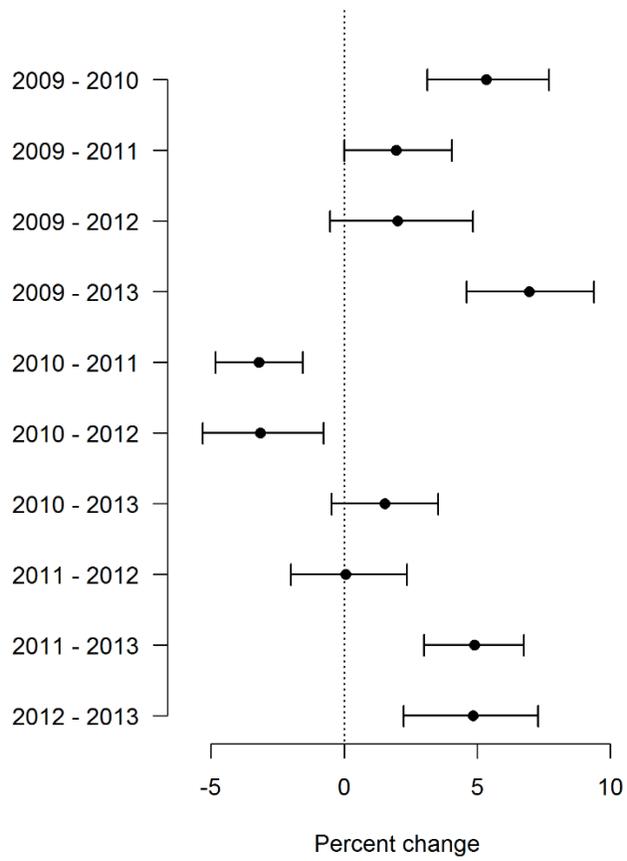

Figure 4. Pairwise estimates of percent change among years in the mean adult body size of *P. shermani*.

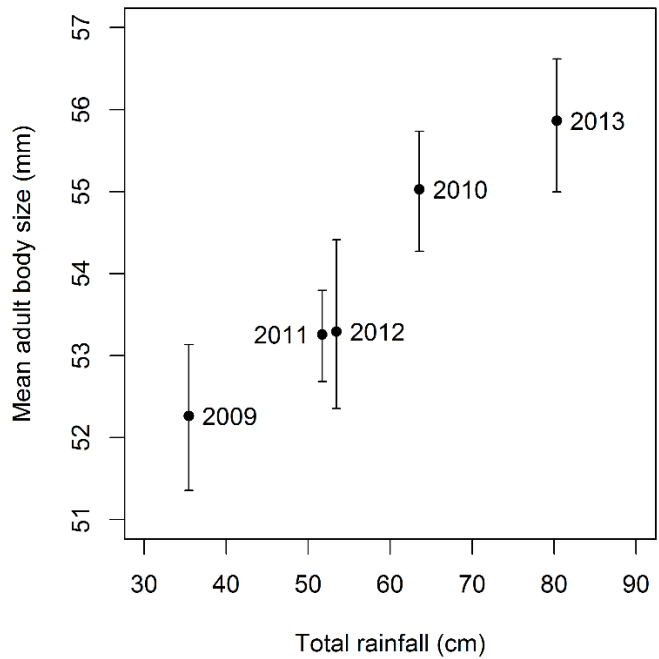

Figure 5. Estimated mean adult body size of *P. shermani* based on a normal mixture model. Error bars indicate 95% credible intervals (CRI) for each annual estimate. Annual rainfall totals are calculated from May 15 – August 15, representing the typical survey window during our study.

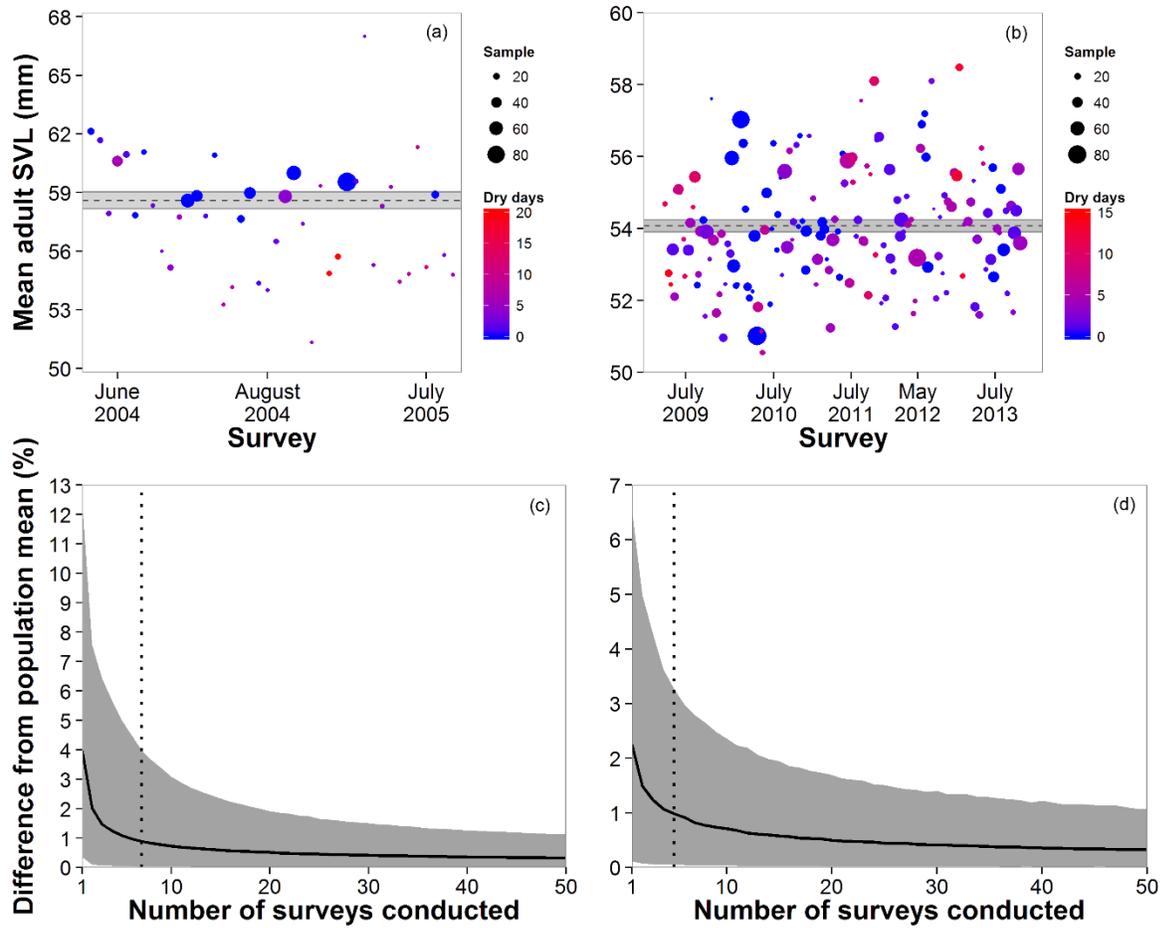

Figure 6. Top panels represent survey means of adult body size for *P. metcalfi* (a), and *P. shermani* (b) with the gray shaded region indicating a 95% confidence band around the overall mean based on all sampled individuals. Bottom panels indicate the average bias (black line) of mean adult body size calculated from randomly selected surveys of *P. metcalfi* (c) and *P. shermani* (d). The gray shaded area indicates the 95% confidence region and the vertical dashed line denotes where the difference between the sample mean and the overall population mean is < 1%.

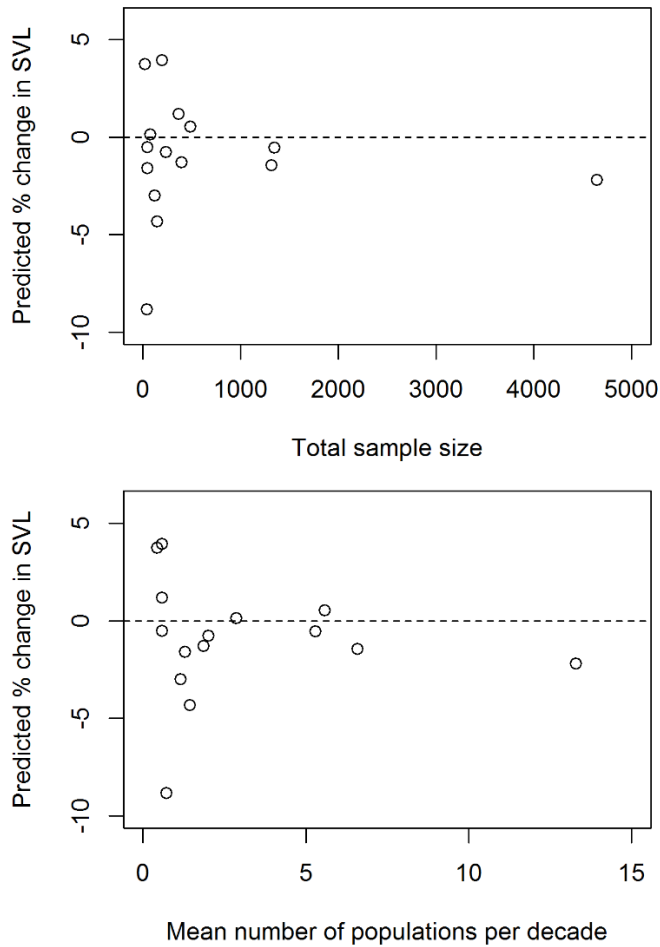

Figure 7. Data from Table 1 of Caruso et al. (2014). Relationship between sampling effort and predicted percent change in snout-vent length (SVL) by species. The top panel represents sampling effort as the total number of individuals measured while the bottom panel represents sampling effort as the mean number of populations surveyed per decade. Predicted percent change over 55 years was calculated using the estimated annual change in SVL (reported "Slope" values from species-specific linear regression or linear mixed effects models) and the average SVL from the first decade of sampling as a reference point.

**Appendix S1.** *A priori* models considered as predictors of mean adult body size (SVL) and frequency in linear mixed models

| Model Name | Parameters |
|---|---|
| Global | Days since soaking rain, Julian day, Julian day$^2$ |
| Rain + Date | Days since soaking rain, Julian day |
| Rain | Days since soaking rain |
| Date$^2$ | Julian day, Julian day$^2$ |
| Date | Julian day |
| Null | Intercept only |

**Appendix S2:** Additional mixture analysis details, prior specification and R code.

We fit the normal mixture model using Markov chain Monte Carlo (MCMC) simulation with the program JAGS (Plummer 2003), executed using the R2jags package (Su and Yajima 2014) in program R (R Development Core Team 2012). We assigned uninformative priors to all model parameters. The mixture weights, $\omega_c$, were assumed to follow a symmetric Dirichlet prior (1,1,1) and gamma priors were assigned ($1.0^{-4}$, $1.0^{-4}$) to the precision of each normal distribution ($1/\sqrt{\sigma_c}$). We chose a normal (0, $1.0^{-6}$) prior for $\mu_1$ and calculated $\mu_2$ as ($\mu_1 + \theta_1, where\ \theta_1 > 0$). Similarly, $\mu_3$ was calculated as ($\mu_2 + \theta_2, where\ \theta_2 > 0$). This parameterization ensures that body size observations are appropriately divided among the three distinct components of the mixture when vague priors are used to define the distribution means. The parameters $\theta_1$ and $\theta_2$ were assigned half-normal priors (0, $1.0^{-6}$)I(0,∞). We generated posterior summaries from 200,000 MCMC iterations of three parallel chains that we thinned 1/20 following a burn-in of 100,000 iterations. This resulted in adequate convergence of all model parameters ($\hat{R}$<1.02; Gelman and Hill, 2007).

```
##########################################################################################
# Bayesian Mixture Analysis
# 16 January, 2014
# Grant M. Connette (grmcco@gmail.com)
# University of Missouri
##########################################################################################

# Libraries
require(mixtools)
require(R2jags)

# Simulate Some Data (based on P. shermani results)
set.seed(101)
NIND <- 1000     # sample size
prob1 <- .23     # probability that an individual is in the 1st age/size class
prob2 <- .39     # probability that an individual is in the 2nd age/size class
prob3 <- .38     # probability that an individual is in the 3rd age/size class
meansizes <- c(24,40,55) # mean size for each class
sdsizes <- c(3.4,4.4,5.4) # sd for each distribution

y1 <- rnorm(n=prob1*NIND,mean=meansizes[1],sd=sdsizes[1]) # random sizes of age 1 individuals
y2 <- rnorm(n=prob2*NIND,mean=meansizes[2],sd=sdsizes[2]) # random sizes of age 2 individuals
y3 <- rnorm(n=prob3*NIND,mean=meansizes[3],sd=sdsizes[3]) # random sizes of age 3 individuals
y <- c(y1,y2,y3)

hist(y,breaks=30)  # Examine simulated size distribution

# Mixture analysis using Mixtools
mixmod <- normalmixEM(y,k=3)
(PROBS <- mixmod$lambda)
(SDSIZES <- mixmod$sigma)
(MEANSIZES <- mixmod$mu)

plot(mixmod,which=2)
lines(density(y), lty=2, lwd=2)

##########################################################################################
# Bayesian Analysis of the Normal Mixture Model (No fixed Year Effects)
##########################################################################################
sink("size.txt")
cat("
  model {

  # Priors for mean size of each class
  mu[1] ~ dnorm(25,1.0E-4)
  mu[2] <- mu[1]+theta1
  theta1 ~ dnorm(0,1.0E-4)I(0,)
  mu[3] <- mu[2]+theta2
  theta2 ~ dnorm(0,1.0E-4)I(0,)

  # Assign priors to standard deviation for each of the size classes
  for (i in 1:3){
  tau[i] ~ dgamma(0.01,0.01)
  sigma[i] <- pow(tau[i],-0.5)
  }

  # Assign priors to the probabilities for membership in each size class (must sum to 1)
  prob[1:3] ~ ddirich(alpha[])
```

```r
    # Alternative prior specification for mixture weights converges in ~ 80% fewer iterations
    #prob[1] ~ dunif(0,1)
    #diff <- 1 - prob[1]
    #prob[2] ~ dunif(0,diff)
    #prob[3] <- 1 - prob[1] - prob[2]

    # Likelihood
    for(i in 1:nind){
    class[i] ~ dcat(prob[]) # choose which distribution to try
    y[i] ~ dnorm(mu[class[i]],tau[class[i]])  # Body size (y) of individual i is normally distributed
    }                              # with a mean and sd dependent on the estimated class
    }
    ",fill = TRUE)
sink()

# Bundle data
win.data <- list(y=y,nind=length(y),alpha=c(1,1,1))

# Initial values
inits <- function(){list(mu=c(runif(1,15,40),NA,NA),
                theta1=runif(1,10,20),
                theta2=runif(1,10,20),
                tau=runif(3,0.03,0.1),
                #prob=c(0.2,0.4,NA),  # if using alternate prior
                class=sample(x=c(1,2,3),size=length(y),replace=T))}

# Parameters monitored
params <- c("mu","sigma","prob")

# MCMC settings
ni <- 10000
nt <- 10
nb <- 5000
nc <- 3

# Call JAGS from R (~ 1 min)
system.time(OUT <- jags(win.data, inits, params, "size.txt", n.chains = nc, n.thin = nt, n.iter = ni, n.burnin = nb))

# Or use WinBUGS
# library(R2WinBUGS)
# system.time(OUT <- bugs(win.data, inits, params, "size.txt", n.chains = nc, n.thin = nt, n.iter = ni, n.burnin = nb, debug=TRUE))

# Summarize posteriors
print(OUT, dig = 4)
OUT.MUs <- OUT$BUGSoutput$mean$mu
OUT.SDs <- OUT$BUGSoutput$mean$sigma
OUT.PROBs <- OUT$BUGSoutput$mean$prob

# Table comparing truth and model estimates
Table <- data.frame(rbind(c(meansizes,sdsizes,c(prob1,prob2,prob3)),c(OUT.MUs,OUT.SDs,OUT.PROBs)))
row.names(Table) <- c("True","Estimated")
names(Table) <- c("mu[1]","mu[2]","mu[3]","sd[1]","sd[2]","sd[3]","prob[1]","prob[2]","prob[3]")
print(Table)
```

```
#################################################################################################
# Bayesian Analysis of Normal Mixture Model with Fixed Year Effects on mu, sigma, prob
#################################################################################################
sink("size.txt")
cat("
model {

# Priors for mean size of each class (differing by year)
for (t in 1:5){
  mu[1,t] ~ dnorm(25,1.0E-6)
  mu[2,t] <- mu[1,t]+theta1[t]
  theta1[t] ~ dnorm(0,1.0E-6)I(0,)
  mu[3,t] <- mu[2,t]+theta2[t]
  theta2[t] ~ dnorm(0,1.0E-6)I(0,)

# Assign priors to standard deviation for each of the size classes (differing by year)
  for (c in 1:3){
    tau[c,t] ~ dgamma(0.0001,0.0001)
    sigma[c,t] <- pow(tau[c,t],-0.5)
  } # close c-loop (class)

# Assign priors to the probabilities for membership in each size class
  prob[1:3,t] ~ ddirich(alpha[,t])
} # close t-loop (year)

# Likelihood
for(i in 1:nind){
  class[i] ~ dcat(prob[,year[i]]) # choose which distribution to try
  y[i] ~ dnorm(mu[class[i],year[i]],tau[class[i],year[i]]) # i.e. body size, y, of individual i is normally distributed
}                          # with a mean and sd dependent on the class membership (which is estimated)

# Derived Quantities - Percent change in mean adult body size between years
muC[1]<-(mu[3,2]-mu[3,1])/mu[3,1]     # 2009-2010
muC[2]<-(mu[3,3]-mu[3,1])/mu[3,1]     # 2009-2011
muC[3]<-(mu[3,4]-mu[3,1])/mu[3,1]     # 2009-2012
muC[4]<-(mu[3,5]-mu[3,1])/mu[3,1]     # 2009-2013
muC[5]<-(mu[3,3]-mu[3,2])/mu[3,2]     # 2010-2011
muC[6]<-(mu[3,4]-mu[3,2])/mu[3,2]     # 2010-2012
muC[7]<-(mu[3,5]-mu[3,2])/mu[3,2]     # 2010-2013
muC[8]<-(mu[3,4]-mu[3,3])/mu[3,3]     # 2011-2012
muC[9]<-(mu[3,5]-mu[3,3])/mu[3,3]     # 2011-2013
muC[10]<-(mu[3,5]-mu[3,4])/mu[3,4]   # 2012-2013
muC[11]<-(abs(muC[1])+abs(muC[5])+abs(muC[8])+abs(muC[10]))/4  # Average difference among sequential years
}
",fill = TRUE)
sink()

# Bundle data
win.data <- list(y=psherm$SVL,nind=length(psherm$SVL),year=psherm$YearID,alpha=matrix(1,nrow=3,ncol=5))

# Initial values
inits <- function(){list(mu=matrix(c(runif(5,15,35),rep(NA,times=10)),nrow=3,byrow=T),
            theta1=runif(5,10,20),
            theta2=runif(5,10,20),
            tau=matrix(runif(15,0.03,0.2),nrow=3),
            class=sample(x=c(1,2,3),size=length(psherm$SVL),replace=T))}

# Parameters monitored
params <- c("mu","sigma","muC")
```

```
# MCMC settings
ni <- 200000
nt <- 20
nb <- 100000
nc <- 3

# Call WinBUGS from R (BRT 3 min)
system.time(SizeOUT <- jags(win.data, inits, params, "size.txt", n.chains = nc, n.thin = nt, n.iter = ni, n.burnin = nb))

# Summarize posteriors
print(SizeOUT, dig = 4)
```